\documentclass{article}

\usepackage{PRIMEarxiv}

\usepackage[utf8]{inputenc} 
\usepackage[T1]{fontenc}    
\usepackage{hyperref}       
\usepackage{url}            
\usepackage{booktabs}       
\usepackage{amsfonts}       
\usepackage{nicefrac}       
\usepackage{microtype}      
\usepackage{lipsum}
\usepackage{amsthm}
\usepackage{fancyhdr}       
\usepackage{graphicx}       
\graphicspath{{media/}}     

\usepackage{amsmath,amssymb,amsfonts}
\usepackage{algorithmic}
\usepackage{graphicx}
\usepackage{textcomp}
\usepackage{multirow}
\usepackage{array}
\usepackage{subfigure}
\usepackage{amsmath,nccmath}
\usepackage{amsfonts}
\usepackage{tablefootnote}

\newcommand{\probP}{\text{I\kern-0.15em P}}

\newtheorem{theorem}{Theorem}[section]
\newtheorem{corollary}[theorem]{Corollary}
\newtheorem{preposition}[theorem]{Proposition}
\newtheorem{assumption}{Assumption}
\newtheorem{lemma}[theorem]{Lemma}
\newtheorem{definition}[theorem]{Definition}

\newtheorem{property}{Property}
\newcommand{\figref}[1]{Fig.~\ref{#1}}

\title{The epistemic dimension of algorithmic fairness: assessing its impact in innovation diffusion and fair policy making}

\author{
  Eugenia Villa\\
  Politecnico di Milano \\
  Milan, Italy\\
  \texttt{eugenia.villa@polimi.it} \\
  \And
  Camilla Quaresmini \\
  Politecnico di Milano \\
  Milan, Italy\\
  \texttt{camilla.quaresmini@polimi.it} \\
  \And
  Valentina Breschi \\
  Eindhoven University of Technology \\
  Eindhoven, The Netherlands\\
  \texttt{v.breschi@tue.nl} \\
    \AND
   Viola Schiaffonati \\
   Politecnico di Milano \\
   Milan, Italy \\
   \texttt{viola.schiaffonati@polimi.it} \\
   \And
   Mara Tanelli \\
   Politecnico di Milano \\
   Milan, Italy \\
   \texttt{mara.tanelli@polimi.it} \\
}

\begin{document}
\maketitle

\begin{abstract}
Algorithmic fairness is an expanding field that addresses a range of discrimination issues associated with algorithmic processes. However, most works in the literature focus on analyzing it only from an ethical perspective, focusing on moral principles and values that should be considered in the design and evaluation of algorithms, while disregarding the epistemic dimension related to knowledge transmission and validation. However, this aspect of algorithmic fairness should also be included in the debate, as it is crucial to introduce a specific type of harm: an individual may be systematically excluded from the dissemination of knowledge due to the attribution of a credibility deficit/excess. In this work, we specifically focus on characterizing and analyzing the impact of this credibility deficit or excess on the diffusion of innovations on a societal scale, a phenomenon driven by individual attitudes and social interactions, and also by the strength of mutual connections. Indeed, discrimination might shape the latter, ultimately modifying how innovations spread within the network. In this light, to incorporate, also from a formal point of view, the epistemic dimension in innovation diffusion models becomes paramount, especially if these models are intended to support fair policy design. For these reasons, we formalize the epistemic properties of a social environment, by extending the well-established Linear Threshold Model (LTM) in an epistemic direction to show the impact of epistemic biases in innovation diffusion. Focusing on the impact of epistemic bias in both open-loop and closed-loop scenarios featuring optimal fostering policies, our results shed light on the pivotal role the epistemic dimension might have in the debate of algorithmic fairness in decision-making.
\end{abstract}


\section{Introduction}
By addressing a range of discrimination issues associated with algorithmic processes \cite{dolata2022,nachbar2020algorithmic}, algorithmic fairness is becoming increasingly central in debates on automated decision-making. This is especially strong in the context of Machine Learning (ML) \cite{weerts,wang,mitchell,caton,wan,pessach,alves}, where algorithmic unfairness is characterized as an individual \cite{dwork,gummadi,bellamy2018ai,biasmitigationaif,kusner2018counterfactual} or a group \cite{biasmitigationaif,hardt,mehrabi2022survey} problem. However, algorithmic fairness is relevant beyond ML \cite{kleinberg2018algorithmic}, and should also be at the center of the debate when considering other approaches to automatic decision making, e.g., control and automation techniques \cite{rao1988}, typically used in engineering fields and yet potentially adoptable for policy design on a societal scale~\cite{breschi2024}.

While there are multiple ways to conceptualize fairness in algorithmic contexts, connected for example to predictive (e.g., \cite{lum2016statistical,quaresmini2023data}) or allocative (e.g., \cite{elzayn2019fair,quaresmini2023qualification}) issues \cite{beigang,fabris}, algorithmic fairness is typically analyzed by adopting an ethical perspective, focusing on moral principles and values that should be considered in the design and evaluation of algorithms \cite{binns2018,lee2021}. In this work, we claim that also the epistemic dimension, related to knowledge transmission and validation, should play a central role in this debate. Without considering this epistemic dimension, there might be the risk of not including in the current debate a specific type of harm \cite{battaglia2024}: an individual may be systematically excluded from knowledge dissemination because others do not perceive them as sufficiently credible. This is the core concept of the epistemic injustice theory proposed by Miranda Fricker \cite{Fricker2007-FRIEIP}. According to this theory, social features have an impact on the agents' credibility, i.e., on the ability to influence their peers \cite{Lackey2018-LACCAT,tian2021social,wang2024social}. This can lead to discrimination in a way in which someone can be systematically attributed a \textit{credibility deficit} and consequently excluded from information sharing.
Although some attempts to analyze the epistemic dimension of fairness \cite{kim,medvecky,miller} in terms of epistemic injustice \cite{jalali,kim,medvecky} can be found in the literature, the idea of extending algorithmic fairness in an epistemic direction has not been extensively discussed so far. However, it is important to address a form of injustice, that is, epistemic injustice, which is not usually addressed by common fairness strategies \cite{edenberg}.
To show how the epistemic dimensions and related harms impact automated decision-making in a dynamic context, we consider the problem of formally characterizing innovation diffusion over a social network. In this context, social interactions play a major role, shaping one's predisposition to embrace the adoption of new technologies \cite{jackson2008social,ravazzi2021learning}. This feature makes networked systems an ideal framework for describing and analyzing processes like adoption dynamics within innovation diffusion, see, e.g., \cite{delre2010will,acemoglu2011diffusion}, while analyzing the impact of existing credibility deficits.

Among existing opinion dynamics models, the deterministic Linear Threshold Model (LTM) 
\cite{granovetter1978threshold} is widely used (e.g., \cite{jackson2006diffusion, beaman2021can,breschi2022fostering}) as
it allows one to account for two key aspects that guide innovation diffusion, that is, personal attitudes and the influence of neighboring individuals, while remaining rather simple. Indeed, given a set of individuals immersed in a network (\emph{agents}), this model describes adoption dynamics as a cascading phenomenon driven by the relative popularity of the new technology among neighbors and by individual attitudes, dictating the impact that such popularity has on one's adoption decision. The LTM has been extended to account for inherent complexities in innovation diffusion processes, e.g., competitive influences \cite{borodin2010threshold}, multiplex networks \cite{yaugan2012analysis}, and the presence of stochastic components driving opinion dynamics \cite{shakarian2015independent}. However, these models do not yet account for possible differences in the strength of mutual bonds among individuals, which in reality affect the diffusion process \cite{van2011opinion}.
To cope with this limitation, other works have adapted the LTM to account for the heterogeneous strength of interactions between individuals \cite{kempe2003maximizing,cox2017spread}. These extensions generally incorporate the latter in the standard LTM by considering weighted networks where the edges' weights describe the individual capability of influencing one's neighbors. However, these works directly focus on 
influence maximization problems, falling short in describing how this unbalanced relation among individuals affects the adoption dynamics in the first place. An attempt in this direction is made in \cite{unicomb2018threshold}, where the impact of the weights' distribution on the time of cascade emergence is analyzed. Yet, this study solely examines the influence of varying connection weights in the absence of external interventions, overlooking the impact of unbiased relations in shaping fostering policies and, eventually, causing marginalization and discrimination among individuals when they are enacted.

By investigating how knowledge is acquired, validated, and disseminated within society, theories from social epistemology can become valuable assets in understanding and, hence, modeling the diffusion of innovative ideas and technologies at a societal scale (as proven by some recent studies, e.g., \cite{moldoveanu2021epistemic,Spiekermann2019-SPIENI,jalali}). Meanwhile, they can provide useful insights into the epistemic dimension, that is related to knowledge transmission and validation, of the assumptions underlying existing innovation diffusion models. Extending the analysis of such dynamics in an epistemic direction allows us to identify a distinct type of harm, leading to epistemic injustice, that otherwise remains ignored. In this light, to incorporate, also from a formal point of view, the epistemic dimension in innovation diffusion models becomes paramount, especially if these models are intended to aid policy design. 
In this work, we thus propose an extension of the LTM \cite{granovetter1978threshold} that explicitly accounts for the \emph{epistemic dimension} and its influence on individuals in driving innovation diffusion. 
Inspired by Miranda Fricker's work \cite{Fricker2007-FRIEIP}, we formalize the concept of \emph{ epistemic fairness in the context of innovation diffusion and fostering policy design}, extending the current discussion on algorithmic fairness in an epistemic direction. By focusing on the lack of epistemic fairness and its impact, first in open-loop and then in a closed-loop scenario featuring optimal fostering policies, our results shed light on the pivotal role the epistemic dimension might have in the debate of algorithmic fairness with a focus on policy design. 




The work is organized as follows. In Section \ref{sec:motivation}, we provide a motivating example showcasing the importance of considering the epistemic dimension in processes of innovation diffusion. We then provide the preliminaries on the standard LTM followed by its epistemic extension in Section \ref{sec:preliminaries}. The concept of epistemic fairness in a social network is then formalized in Section \ref{sec:epistemic}, allowing us to analyze the impact of epistemic unfairness in free and forced adoption dynamics (see Sections \ref{sec:open_loop}-\ref{sec:closed_loop}). 
The work terminates with some final remarks and directions for future work.


\section{Motivating Example: the Quest to Achieve Sustainability Goals}\label{sec:motivation}
The process of adopting innovative sustainable technologies \cite{sinha2022analyzing,tarekegne2020just} (e.g., sharing mobility, photovoltaic energy, electric vehicles (EV)) is deemed key for a substantial reduction in environmental impact and harmful emissions worldwide. 
At the same time, individual attitudes, resistance to change, and societal inertia \cite{ rogers2014diffusion,umar2022role} play a key role in impeding the widespread adoption of new technologies. These challenges could be addressed by pairing governmental incentive policies (see, e.g., \cite{eu_green,US_green}) with a robust social structure that encourages diffusion through imitation \cite{nejad2014influentials}, with early adopters persuading hesitant individuals by demonstrating adoption-related benefits.
However, different types of social bias can hinder the formation of such a social structure, diminishing the positive impact of imitation and slowing down the diffusion of innovation. Hence, accounting for these biases at policy design time and understanding their impact on the effectiveness of deployed fostering policies becomes crucial for these interventions to be ultimately effective. 

In light of these considerations and with a focus on epistemic biases, i.e. prejudices in the production, validation and share of knowledge,
this work formalizes the concept of \emph{algorithmic} epistemic fairness in innovation diffusion in order to incorporate it into policy design. By integrating epistemic considerations with the technical tools of control theory, our approach can become a tool for the fair dissemination of sustainable technologies, driven by social contagion and achieved through the reduction of epistemic bias.



\section{An Epistemically Grounded Model for Innovation Diffusion}\label{sec:preliminaries}
Our starting point 
is the well-known LTM introduced in \cite{granovetter1978threshold}. By modeling the adoption choice as a binary variable associated with each individual, this model describes the adoption dynamics as a deterministic combination of personal attitude and influence from neighbors, considering a set of individuals immersed in a network.

Let us model a social environment through an undirected, strongly connected graph $\mathcal{G}=(\mathcal{V},\mathcal{E})$, where the set of nodes $\mathcal{V}$ represents the individuals within a social community, and the set of edges $\mathcal{E}$ indicates the existence of a social connection between two agents. Accordingly, given $x,y \in \mathcal{V}$, the following holds

\begin{equation}\label{eq:neighbours}
    y \in N_{x} \iff (x,y) \in \mathcal{E},
\end{equation}

where $N_{x}$ denotes the set of neighbors of node $x$, i.e., the set of individuals that can influence the agent's opinion. Note that since the graph is strongly connected, there are no isolated communities, and a path always exists between all pairs of agents $x$ and $y$ in $\mathcal{V}$. Consequently, all agents influence (directly or indirectly) each other. Each $x \in \mathcal{V}$ is paired with a binary variable $a_x(t) \in \{0,1\}$ denoting the agent's adoption state at time $t \in \mathbb{N}$, with $a_x(t)=1$ if agent $x \in \mathcal{V}$ is an adopter of the considered technology at time $t$ and zero otherwise. By relying on the agents' states, we introduce the following assumption \cite{granovetter1978threshold}.
\begin{assumption}[Seed set]\label{assump:1}
    There exists a \emph{non-empty} set of agents (the \emph{seed set})  $S^{\star}(0) \neq \emptyset$ such that
    \begin{equation}\label{eq:seed_set}
        S^{\star}(0)=\{x \in \mathcal{V} | a_{x}(0)=1\}.
    \end{equation}
\end{assumption}
Along a similar line of the definition of $S^{\star}(0)$, we introduce the set of agents that have embraced the new technology at time $t \in \mathbb{N}$ as follows:
\begin{equation}\label{eq:adopter_set}
    S^{\star}(t)=\{x \in \mathcal{V}|a_{x}(t)=1\},
\end{equation}
defining the (irreversible) innovation diffusion dynamics as
\begin{equation}\label{eq:LTM}
    a_{x}(t+1)=
    \begin{cases}
        1, & \mbox{if } a_{x}(t)=1 \lor \dfrac{\left|N_{x}\cap S^{\star}(t)\right|}{|N_{x}|}\geq \rho_{x},\\[1mm]
        0, & \mbox{otherwise,}
    \end{cases}\quad \forall t \in \mathbb{N},\; x \in \mathcal{V}.
\end{equation}
Note that, according to this model, once agents become adopters, they can no longer change their status. Meanwhile, their inclination toward adoption is driven by the \emph{individual resistivity} $\rho_{x} \in [0,1]$, for $x \in \mathcal{V}$.
At each time step, synchronous communications enable agents to update their adoption state based on the current 
status of their neighbors. Hence, the more individuals adopt the innovation, 
the easier it becomes to persuade others to do the same. This triggers a \emph{contagion effect}, propagating the innovation throughout the network 
\cite{young2009innovation}
.

While being largely used for its simple yet effective representation of the adoption mechanism \cite{jackson2006diffusion, altarelli2013optimizing, beaman2021can,breschi2022fostering}, the LTM presents some clear limitations, especially in 
view of real-world validation. First, 
adopting an innovation is assumed to be irreversible. As individuals can change their views after trying new technologies, this is unrealistic in the long run but justifiable if the LTM is used to model innovation diffusion over a limited period. For instance, significant investments in durable goods (e.g., EV purchases) are often irreversible when considered over a short-term period (e.g., the average lifespan of an EV or its payback period of approximately 10 years). Moreover, the dynamics in \eqref{eq:LTM} is purely deterministic, hence excluding the presence of stochastic elements guiding individual decisions. Although incorporating a stochastic component is crucial for a more realistic depiction of the diffusion process, in this work we maintain a deterministic framework to have closed-form measures of the epistemic impact on innovation diffusion in both open and closed-loop scenarios, which pave the way for their future analysis in a stochastic context.

One of the core not generally emphasized assumptions of the model in \eqref{eq:LTM} is that individuals implicitly have the same capacity to influence their neighbors, neglecting the epistemic properties of the agents. Nonetheless, what we call \emph{epistemic power} - that is, the capacity to influence others - is generally not equally shared among individuals. On the contrary, some agents have a greater capacity to influence their peers \cite{van2011opinion} and disregarding this dimension in innovation diffusion models might lead to a less realistic representation of the adoption mechanisms. Furthermore, ignoring the epistemic dimension in modeling innovation diffusion mechanisms leads to the impossibility of grasping a specific type of injustice, namely \emph{epistemic injustice} \cite{Fricker2007-FRIEIP}, which nonetheless may arise in social relations. The core idea of this influential theory by Miranda Fricker is that agents have an epistemic power directly connected to the social features they are characterized by. For instance, a police man might not believe a dark-skinned individual due to to their ethnicity \cite{Fricker2007-FRIEIP}. This power to influence others is expressed by agents' credibility, conceived as how much a speaker is believed by a hearer. In light of empirical evidence that emphasizes the importance of reciprocal aspects in epistemic analyses (see \cite{mahmoodi2018reciprocity}), we formalize this concept as follows.
\begin{definition}[Relational credibility]\label{def:gamma}
    The level of credibility agent $y \in \mathcal{V}$ attributes to agent $x \in \mathcal{V}$ is encoded in a parameter $\gamma_{x,y} \in [0,1]$.
\end{definition}

According to this definition, $\gamma_{x,y} \in [0,1]$ ultimately indicates how much represents how much agent $x$ is believed by agent $y$. This additional parameter allows us to
shape our description of individual adoption propensity as a combination of personal attitude, influence from neighboring agents (dictated by the topology of the social connections), and the epistemic characterization of the environment (driven by the epistemic power associated with each pair of agents). Formally, this translates into the following \emph{epistemic-based} extension of \eqref{eq:LTM}:
\begin{equation}\label{eq:cascade}
    a_x(t+1)=\begin{cases}
        1 \mbox { if } a_x(t)\!=\!1 \!\lor \frac{\sum_{y \in N_x^{\ast}(t)}\!{\gamma_{y,x}}}{\sum_{y \in N_x}\!{\gamma_{y,x}}}\!\geq\!\rho_x\\
        0 \mbox{ otherwise}
    \end{cases}\!t \in \mathbb{N},~x,y \!\in\! \mathcal{V}. 
\end{equation}
In this way, agents become adopters if the influence of their adopting neighbors, weighted by their relational credibility values, exceeds their resistivity threshold $\{\rho_x\}_{x \in \mathcal{V}}$. 

\section{Epistemic Fairness in innovation diffusion}\label{sec:epistemic}

Based on Fricker's theory \cite{Fricker2007-FRIEIP}, when epistemic injustice occurs a speaker is not able to share their knowledge with a hearer, due to the attribution of a credibility deficit (or excess) based on certain social features. Building on this, we could distinguish two types of epistemic biases: societal and relational. To give an example of the former, consider a dark-skinned female engineer who is highly reliable on providing suggestions on new technologies, however having a reduced power of influence due to her being a dark-skinned woman in a biased society (tending to privilege white men). Instead, in cases where the hearer shares some features with the speaker (hence being more likely to take their advice according to the homophily principle \cite{tang2013}), we can still face a relational epistemic bias, i.e., a power relationship that scales the societal bias.
Formally, this translate in the possibility of agents to have a credibility deficit (or excess). To conceptualize this possibility, we introduce the notion of reliability.
\begin{definition}[Individual reliability]\label{def:r}
    Reliability is a characteristic $r_x$ of agent $x \in \mathcal{V}$, representing how much agent $x$ should be believed by any other agent in $\mathcal{V}$.
\end{definition}
Although not explicitly introduced in \cite{Fricker2007-FRIEIP}, reliability is here considered to interpret the condition for which subjects who suffer epistemic injustice are perceived as less (or more) credible than they ought to be. Based on this definition, reliability is an individual attribute, which can be determined by proxies to it,  such as educational degree, specialized qualifications, or the level of experience, for example in a work environment. Reliability is not influenced by epistemic biases, hence representing the potential credibility of an agent in the absence of such prejudices.
While without prejudice $\gamma_{x,y}\!=\!r_{x}\in [0,1]$ for all $x,y \!\in\! \mathcal{V}$, existing biases that drive discrimination connect reliability to credibility by attributing some agents a credibility deficit/excess $\Delta_{x,y} \in [r_x-1,r_x]$ based on the agent's sensitive features, with agents $x$ having a credibility deficit if $\Delta_{x,y} \in (0,r_x]$ and a credibility excess\footnote{Saying that agent $x \in \mathcal{V}$ has a credibility deficit makes sense only if $r_x \neq 0$, while a credibility excess is reasonable only when $r_x \neq 1$.} when $\Delta_{x,y} \in [r_x-1,0)$  
. The credibility deficit and excess are determined by a biased perception of specific social characteristics that reduces or augment the agents' ability to be believed by peers \cite{fricker2}. Therefore, in the presence of societal epistemic bias, agents' reliability 
becomes their credibility, that is, $\gamma_{x,y} = r_x - \Delta_{x,y},~~x,y \in \mathcal{V}$ if an individual is subject to a credibility deficit.


Toward quantifying relational credibility according to the previous definition, we build upon the individual credibility dimension introduced in \cite{fPET23}. According to the latter, societal epistemic bias can be evaluated using an intersectional discrimination factor $\phi_x \in [0,1]$ for all $x \in \mathcal{V}$, whose value increases if an individual belongs to one or more marginalized groups. This factor is defined as $\phi_x = 0.5^{\sum_{g} d_x^g}$, where $d_x^g = 1$ if $x$ belongs to the discriminated group $g$ and zero otherwise, and it changes the individual resistivity, leading to the individual credibility as
\begin{equation}
    \gamma_x=\phi_x r_x,~~\forall x \in \mathcal{V}. 
\end{equation}
Therefore, if individuals do not belong to any discriminated group, their credibility and reliability coincide.
We then consider that relational epistemic biases can exacerbate (or mitigate) such an individual credibility deficit, that is attributed at the societal level. This concept is formalized as follows:
\begin{equation}
    \gamma_{x,y}=\phi_{x}\eta_{x,y}r_x,~~\forall x,y \in \mathcal{V},
\end{equation}
where $\eta_{x,y} \in [0,1]$ indicates how much the reliability of agent $x$ (already modified by the individual discrimination factor $\phi_x$) is affected by a relational form of discrimination by agent $y$. This relational discrimination factor is defined similarly to $\phi_x$ as
\begin{equation}
    \eta_{x,y}=0.5^{\pm\sum_g d_x^{g}d_y^{g}},~~ x,y \in \mathcal{V}, 
\end{equation}
with $\eta_{x,y}=0.5^{\sum_g d_x^{g}d_y^{g}}$ characterizing a credibility deficit and $\eta_{x,y}=0.5^{-\sum_g d_x^{g}d_y^{g}}$ a credibility excess. Indeed, when $x$ and $y$ belong to the same minority group, this mitigates the effect of the individual discrimination factor, ultimately removing the discrimination for same-group individuals. Note that, while the relational discrimination factor is symmetric, $\gamma_{x,y} \neq \gamma_{y,x}$ due to the individual credibility and, thus, relational credibility is overall asymmetric.

Based on such epistemic characterization, we define \emph{epistemic fairness} as follows.
\begin{definition}[Epistemic Fairness]\label{EFair} 
A social network $\mathcal{G}$ is said epistemically fair, denoted as EF($\mathcal{G}$), if and only if $\gamma_{x,y}=r_x,~~\forall x,y \in \mathcal{V}$.
\end{definition}

This definition implies that all agents $x,y \in \mathcal{V}$ in an epistemically fair network have an amount of credibility $\gamma_{x,y}$ sufficient not to be wronged in their capacity as knowers \cite{Fricker2007-FRIEIP}, i.e., they are attributed the amount of credibility they would have in the absence of epistemic bias.
This implies that epistemic fairness is attained when agents' epistemic power corresponds to that of the ideal unbiased scenario, namely

\begin{equation}\label{eq:epsitemi_fairness_when}
    \Delta_{x,y}=0, ~~\forall x,y \in \mathcal{V}.
\end{equation}

Therefore, epistemic injustice occurs when someone is assigned a credibility deficit or excess due to sensitive features, e.g., age, gender, ethnicity, etc. This leads to the unfair treatment of such individuals regarding their ability to access and contribute to knowledge production.\footnote{We acknowledge the complexity of the epistemic injustice theory. Here, we reduce the concept to its \emph{testimonial} pattern (i.e., when someone is downgraded in their credibility due to sensitive features), leaving aside the \emph{hermeneutical} one (i.e., when someone's experience is ignored, for the same reason), as it is more relevant to our work.} Note that, along the lines of \cite{Fricker2007-FRIEIP}, the condition in~\eqref{eq:epsitemi_fairness_when} does not presume equal credibility for all the agents, but it rather implies that they are trusted based on their reliability, namely the actual knowledge they possess (on a specific topic), which can still be unequally distributed.

\section{The Impact of Epistemic (Un)Fairness} \label{sec:open_loop}
In the footsteps of \cite{acemoglu2011diffusion}, which investigates the impact of the network topology and individual resistivity on LTM, we analyze the effects of introducing the epistemic dimension in the diffusion process modeled as in \eqref{eq:cascade}. Towards this objective, we extend the definition of cohesive set provided in \cite{acemoglu2011diffusion} as follows.

\begin{definition}[Epistemically cohesive set]\label{def:coehsive_set}
    A set $X\subseteq \mathcal{V}$ is epistemically cohesive if 
    \begin{equation}\label{eq:choesive_set}
        \frac{\sum_{y \in (X \cap N_x)}{\gamma_{y,x}}}{\sum_{y \in  N_x}{\gamma_{y,x}}}>1-\rho_{x},~~\forall x \in X.
    \end{equation}
\end{definition}

Therefore, for a set $X \subseteq \mathcal{V}$ to be epistemically cohesive, the ratio between the credibility attributed to neighboring agents belonging to $X$ and the influence of all the neighboring agents is larger than $1-\rho_x$ for all $x \in X$. 
Consequently, an epistemically cohesive set suffers from \emph{epistemic isolation}, making it inaccessible to transferred knowledge and, therefore, impenetrable with respect to the adoption of new technologies. The asymmetry in relational credibility further reinforces this effect, as different agents may perceive the credibility of the same source differently, influencing the diffusion process in a non-uniform manner. For the sake of our analysis, we also recall the following properties of cohesive sets that extend to our definition of an epistemically cohesive set.
\begin{property}[Empty set]\label{property1}
     The empty set $\emptyset$ is epistemically cohesive.
\end{property}
\begin{property}[Union of sets]\label{property2}
 The union of epistemically cohesive sets is epistemically cohesive.
\end{property}

Moreover, we introduce the set $\mathcal{S}(t)$ of agents switching to the acceptance of the innovation at time $t>0$, i.e.,
\begin{equation}\label{eq:switching_set}
\mathcal{S}(t)=\{x \in \mathcal{V}|a_{x}(t-1)=0 \land a_{x}(t)=1\},
\end{equation}
based on which we introduce the following definition.
\begin{definition}[Fixed point]\label{def:fixed_point}
    The set $\bar{S}$ is a fixed point for the dynamics dictated by \eqref{eq:cascade} if
    \begin{equation}\label{eq:fixed_point}
        S^{\star}(\tau)=\bar{S} \Rightarrow \mathcal{S}(t)=\emptyset, ~~\forall t>\tau\geq0.
    \end{equation}
\end{definition}

To study the impact of the seed set in innovation diffusion, we begin by characterizing the innovation diffusion process at time $t=0$, through the following result.
\begin{preposition}\label{prep:cohesive_set}
    Let $X \in \mathcal{V}$ be an epistemically cohesive set and $X^c=\mathcal{V}\setminus X$ be its complementary. Assume  $S^{\star}(0) \subseteq X^c$. Then, for any $x \in X$ the following holds:
    \begin{equation}\label{eq:choesive_set_influence}
        \frac{\sum_{y \in N_{x}^{\star}(0)} {\gamma_{y,x}}}{\sum_{y \in  N_x}{\gamma_{y,x}}}<\rho_{x},~~~\mbox{ with }~~N_{x}^{\star}(0)=S^{\star}(0)\cap N_x.
    \end{equation}
\end{preposition}
This proposition, whose proof is omitted as it is a direct consequence of Definition~\ref{def:coehsive_set}, formalizes the level of influence of seeds on the agents of an epistemically cohesive set $X$. Indeed, \eqref{eq:choesive_set_influence} states that the influence of neighbors within  $S^{\star}(0)$ is always below the agents' individual thresholds if no early adopters are present in $X$.
Due to the asymmetry of relational credibility, the perceived influence of seeds may vary across different agents in $X$, further reinforcing the epistemic isolation of the set.
This result also implies that none of the agents in a cohesive set $X$ will ever switch to adoption at time $t=0$ if no early adopters are present in $X$ itself (according to~\eqref{eq:cascade}). We then characterize the set $\mathcal{S}(t)$ in \eqref{eq:switching_set} for $t>0$ as follows.
\begin{preposition}\label{prep:fixed_point}
    Given a fixed point $\bar{S}$ such that $S^{\star}(0)=\bar{S}$, then for each $x \in \bar{S}^{c}$ the following holds:
    \begin{equation}\label{eq:corollary341}
    \frac{\sum_{y \in (N_{x} \cap \bar{S})}{\gamma_{y,x}}}{\sum_{y \in N_x}{\gamma_{y,x}}}<\rho_x.
    \end{equation}   
\end{preposition}
Straightfowardly stemming from Definition~\ref{def:fixed_point} and \eqref{eq:cascade}, proposition \ref{prep:fixed_point} implies that once the diffusion process hits a fixed point at time $\tau>0$, the spread of the innovation stops and, hence, no agent in $\bar{S}^c$ will ever switch to adoption after $\tau$. Here the relational characterization of credibility makes an agent’s likelihood of adoption depending not only on the absolute influence of its neighbors but also on how it perceives their credibility, leading to potential asymmetries in the dynamics. Prepositions \ref{prep:cohesive_set}-\ref{prep:fixed_point} suggest a direct relationship between the concept of epistemic cohesiveness and fixed points of the diffusion dynamics, which is now formalized.
\begin{lemma}\label{lemma:1}
    The set $\bar{S}$ is a fixed point for the cascade dynamics described in \eqref{eq:cascade} with $S^{\star}(\tau)=\bar{S}$ if and only if $\bar{S}^c$ is an epistemically cohesive set.
\end{lemma}
Stemming from Definitions~\ref{def:coehsive_set}-\ref{def:fixed_point} and Proposition~\ref{prep:fixed_point}, this result implies that fixed points in adoption depend on how credibility is perceived across agents due to relational credibility. Indeed, if $\bar{S}^c$ is epistemically cohesive, then the influence from outside the set is insufficient to trigger adoption, making $\bar{S}$ a fixed point. The asymmetry in relational credibility further enforces this condition by allowing different agents to assess credibility differently, potentially leading to heterogeneous stopping conditions across the network.
These results jointly allow us to characterize the steady-state behavior of the epistemic-based LTM in \eqref{eq:cascade} with seed set $S^{\star}(0)$.

We now focus on the characterization of the final set $S^{\star\star}= \{x \in \mathcal{V}| a_x(t)=1, t\to +\infty\}$,
which is a fixed point of the epistemic-based LTM dictated by \eqref{eq:cascade}. 
\begin{theorem}[On $S^{\star\star}$]\label{theorem:1}
If there exist \emph{a set of $K>1$} fixed points $\{\bar{S}_{k}\subset\mathcal{V}\}_{k=1}^{K}$ such that
    \begin{equation}
        \left(S^{\star}(0)\subseteq \bar{S}_k \right)\land \left(\bar{S}_k^c \mbox{ is cohesive}\right),~~\forall k\in [K],
    \end{equation}
    then $S^{\star\star} = \mathbb{S}_{K}=\bigcap_{k=1}^{K}{\bar{S}_k}$.
    \begin{corollary}\label{cor:largest_cohesive}
    Let $\mathcal{M}$ be the largest cohesive subset of $\mathcal{V}\setminus S^{\star}(0)$. Then, $S^{\star\star}=\mathcal{M}^c$.
    \end{corollary}\end{theorem}
    The proof follows the same steps of that in~\cite{acemoglu2011diffusion} and it is thus omitted.
To further describe the effect of bias on the \emph{transient} of the diffusion process 
we make the following assumption.
\begin{assumption}\label{assumption:6}
    The set $S^{\star}(t)$ of adopters at time $t\geq 0$ is the same for both the epistemically fair and unfair scenario.
\end{assumption}
Let us then define $N_{x}^{\alpha}$ and $N_{x}^{\beta}$ 
as $N_{x}^{\alpha}=N_{x} \cap S^{\star}(t),~~  N_{x}^{\beta}=N_{x} \cap (S^{\star}(t))^{c},~~t \geq 0$,
and let us introduce the set $S^{\star,\mathrm{fair}}(t+1)$ of adopters (see~\eqref{eq:adopter_set}) at time $t+1$ in the epistemically fair scenario. Then, the following holds.
\begin{preposition}
\label{cor:max_social_cont}
    Under Assumption~\ref{assumption:6}, credibility deficits or excesses do not hinder innovation diffusion at time $t+1$ with respect to an epistemically fair scenario if 
    \begin{equation}\label{eq:condition_transient}
        r_x^{\alpha}\Delta_x^{\beta}\geq r_{x}^{\beta}\Delta_{x}^{\alpha},~~\forall x \in S^{\star,\mathrm{fair}}(t+1)
    \end{equation}
    with $r_x^{\alpha}=\sum_{y \in N_x^{\alpha}}r_{y}$, $\Delta_x^{\alpha}=\sum_{y \in N_x^{\alpha}}\Delta_{y,x}$ and $r_x^{\beta}$ and $\Delta_{x}^{\beta}$ defined in a similar way, by replacing $N_x^{\alpha}$ with $N_x^{\beta}$.
\end{preposition}
\begin{proof}
    The proof can be found in the Appendix.
\end{proof}
This result indicates that an epistemic bias does not necessarily prevent the diffusion of innovation through time within the network with respect to an unbiased context
. At the same time, the individual credibility deficit and excess can have a dual impact depending on its distribution. Indeed, the condition in \eqref{eq:condition_transient} implies that relevant credibility deficits or excesses for positively inclined individuals hinder innovation spread.
This is not the case if the deficit affects non-adopters, as its diffusion can still benefit from the impact of social contagion.

\section{Credibility Deficits and Excesses and Closed-Loop Innovation Diffusion}\label{sec:closed_loop}
Innovation diffusion models explicitly accounting for epistemic biases can be key in supporting the design of fair (resource allocation) policies \cite{quaresmini2023qualification} to nudge virtuous behaviors, e.g., adopting sustainable technologies, using control theoretic tools \cite{roadmap23} and avoiding by-design to exacerbate existing social divides \cite{volodzkiene}. By extending the epistemic-based LTM in \eqref{eq:cascade} to include the effect of controlled inputs and assuming the existence of an external entity (the policymaker) overseeing the network $\mathcal{G}$ and in charge of allocating resources to nudge agents toward virtuous behavior, we 
analyze the impact of 
epistemic fairness on closed-loop adoption dynamics. We focus on the LQR framework \cite{lewis2012optimal} since such a policy design strategy would allow policymakers to achieve conventional goals, such as minimize costs and maximize the impact of a policy, in a structured way. Our analysis assumes the policymaker has \emph{unlimited} resources. 
Although this is a strong idealization, yet it allows us to provide analytical results on the impact of credibility deficits/excesses in closed-loop, and it will be relaxed in future works.

\subsection{A controlled, epistemic-based LTM}
The epistemic-based LTM described by~ \eqref{eq:cascade} is \emph{autonomous}, as it does not feature a dependence on an external input (a policy) that can act on and shape the cascaded mechanism. Toward analyzing the effect of epistemic bias 
in closed-loop, we thus further extended the model to make it non-autonomous. Specifically, in the footsteps of \cite{villa2023fostering}, we make the following assumption.

\begin{assumption}[Policies and adoption dynamics]\label{assump:thrsdyn}
    External (eventually personalized) policies $\{u_{x}(t)\}_{x \in \mathcal{V}}$, with $t \in \mathbb{N}$, solely act on individual resistivity according to
    \begin{equation}\label{eq:dynamic}
        \rho_{x}^{u}(t+1)=\rho_{x}^{u}(t)+b_{x}u_{x}(t),~~\forall x \in \mathcal{V},~t>0,
    \end{equation}
    where $\rho_{x}^{u}(t) \in [0,1]$, $\rho_{x}^{u}(0)=\rho_{x}$ and $\rho_{x}$ is the resistivity featured in \eqref{eq:cascade} and $b_x \in [-1,0)$ for all $x \in \mathcal{V}\setminus S(0)$.
\end{assumption}

Under this assumption, external actions are tailored to modify the impact of adoption barriers on individual choices (and reduce them if $u_{x}(t) \geq 0$, for all $t \in \mathbb{N}$ and all $x\in \mathcal{V}\setminus S^\star(t)$), with an effect that is modulated by $b_{x}$, in turn potentially different for each $x \in \mathcal{V}$. This parameter encapsulates the agent's predisposition to react to a policy, bridging toward a more realistic description of adoption dynamics that considers the unlikeliness of having equal responses to a policy by different individuals. This becomes clear when reconsidering our motivating case (see Section~\ref{sec:motivation}) and focusing on EV adoption. In this case, it is not guaranteed that everyone will take equally advantage of incentives for purchasing an EV if resources are evenly distributed within society (as by generalized price discounts recently promoted by many governments). Indeed, while these economic incentives may be adequate for some people to proceed with an EV purchase, they might only lead to a partial (if any) reduction in resistivity for others, not prompting their adoption of the technology also due to factors other than price (see, e.g., \cite{breschi2023driving,mundaca2020drives}). Note that, based on Assumption \ref{assump:thrsdyn} all agents but the seeds are impacted by the design policy, even if mildly (when $b_x$ is close to zero), guaranteeing that the thresholds' dynamics \eqref{eq:dynamic} is controllable.
Based on \eqref{eq:dynamic}, the dynamics in \eqref{eq:cascade} is thus modified as
\begin{equation}\label{eq:cascade2}
    a_x(t+1)=\begin{cases}
        1 \mbox { if } a_x(t)=1 \lor \frac{\sum_{y \in N_x^{\ast}(t)}{\gamma_{y,x}}}{\sum_{y \in N_x}{\gamma_{y,x}}}\geq\rho_x^{u}(t)\\
        0 \mbox{ otherwise}.
    \end{cases}
\end{equation}

\subsection{Epistemic-based optimal Policy Design}
By relying on \eqref{eq:dynamic}-\eqref{eq:cascade2} and focusing on a finite \emph{horizon} of length $T>0$, we formulate the LQR problem for policy design for the epistemic-based LTM is formalized as follows:
\begin{subequations}\label{eq:optimal_problem}
    \begin{equation}
        \begin{aligned}
            &\underset{\{U_{x}\}_{x \in \mathcal{V}}}{\mathrm{minimize}}~~\sum_{x \in \mathcal{V}} J_{x}(U_{x}) \qquad \mbox{ s.t. }~~\eqref{eq:dynamic}, \eqref{eq:cascade2}, ~~\forall x \in \mathcal{V},
        \end{aligned}
    \end{equation}
    where $U_{x}=\{u_{x}(\tau)\}_{\tau=0}^{T-1}$, the local loss is
    \begin{equation}\label{eq:optimal_cost}
        J_{x}(U_{x})=\sum_{\tau=0}^{T-1}\left[\omega_{x}^{\rho}(t)e_x^2(\tau)+\omega^{u}u_{x}^2(\tau)\right]+\omega_{x}^{\rho}(T)e_x^2(T)
    \end{equation}
    with $\omega_{x}^{\rho}(t),\omega^{u}>0$ are tunable weights calibrating the relative importance of the different terms in the loss and
    \begin{equation}\label{eq:tracking_error}
        e_{x}(\tau)=\rho_{x}^{u}(\tau)-\bar{\rho}_{x},~~\tau=0,\ldots,T,
    \end{equation}
    is the tracking error given the set point $\bar{\rho}_{x}$, $x \in \mathcal{V}$.
\end{subequations}
Besides the terminal cost enforcing error minimization at the end of the horizon, the first two terms in \eqref{eq:optimal_cost} focus on maximizing the boosting effect by reducing individual resistivity towards the target set point $\bar{\rho}_{x}$\footnote{Following \eqref{eq:cascade} reducing individual resistance implies that less influence of neighboring adopters is required for one to adopt the innovation, thereby facilitating its diffusion through a contagion effect.}, while simultaneously minimizing policy efforts. The tunable weights $\{\omega_{x}^{\rho}(t)\}_{x \in \mathcal{V}}$ are here set to $\omega_{x}^{\rho}(t)=\bar{\omega}^{\rho}(1-a_{x}(t)),~~\forall x \in \mathcal{V}$, where $\bar{\omega}^{\rho}>0$ is the actual hyper-parameter to be calibrated.
This choice removes the tracking loss in \eqref{eq:optimal_cost} whenever agents become adopters, pushing the associated inputs to zero from that time onward, thus preventing the waste of resources.
The set point for individual resistance in \eqref{eq:tracking_error} is set to
\begin{equation}\label{eq:individual_target}
\bar{\rho}_{x}=\frac{\sum_{y \in N_{x}^{\star}(0)}{\gamma_{y,x}}}{\sum_{y \in N_x}{\gamma_{y,x}}},~~~\forall x \in \mathcal{V}\setminus S^{\star}(0).
\end{equation}
The latter is the barrier for the cascade effect to impact a non-adopter agent $x$ at time $t=0$, representing the worst-case relative popularity among neighbors of an innovation needed for the agent to accept it. Note that, based on this definition for the target, $\bar{\rho}_{x} \geq 0$ for all $x \in \mathcal{V}\setminus S^{\star}(0)$.\footnote{The calibration of $\bar{\omega}^{\rho}$ and $\omega^u$ is entrusted to policymakers, who can adjust them to balance between conservative actions ($\omega^{u} > \bar{\omega}^{\rho}$) and targeting the effectiveness of promoting strategies ($\bar{\omega}^{\rho} >\omega^{u})$. The detailed analysis on impact of customization will be subject of future research, starting from the analysis carried out in \cite{villa2023fostering} only within epistemically fair networks.}
Along the line of \cite{villa2023fostering}, this problem is solved in a \emph{receding horizon fashion}, i.e., solving at time $t$ the full problem within the finite horizon of length $T$ and then applying only the first optimal action, before proceeding to the next iteration where the same mechanism is applied.
Standard arguments (see e.g. \cite{bertsekas2012dynamic}) can then be used to prove that the control action at each time step $t$ can be parameterized as
\begin{equation}\label{eq:optimal_u}
     u_x(t)=\kappa_x(t) (\rho_x^{u}(t) - \bar{\rho}_x),
\end{equation}
with the personalized policy gain $\kappa_{x}(t) \in \mathbb{R}$ given by $\kappa_x(t)=\tilde{\kappa}_{x}(0|t)$, and $\tilde{\kappa}_{x}(0|t)$ computed backwards through
    \begin{align}\label{eq:controldesign_steps}
        & \tilde{\kappa}_{x}(\tau|t)=-\frac{b_{x}\tilde{p}_{x}(\tau+1|t)}{\omega^{u}+\tilde{p}_{x}(\tau+1|t)b_{x}^2}, \qquad \tilde{p}_{x}(\tau|t)=\omega_{x}^{\rho}(t)+\tilde{p}_{x}(\tau\!+\!1|t)-\frac{b_{x}^2\tilde{p}_{x}^2(\tau\!+\!1|t)}{\omega^{u}+\tilde{p}_{x}(\tau\!+\!1|t)b_{x}^2},
    \end{align}
for $\tau=0,\ldots,T-1$, starting from $\tilde{p}_{x}(T|t)=\omega_{x}^{\rho}(T)$, for all $x \in \mathcal{V}\setminus S^{\star}(t)$. Note that, with this choice of the terminal cost-to-go, the dynamic programming is the exact solution of the 
problem in \eqref{eq:optimal_problem} and, thus, it is unique.

\subsection{Epistemic dimension in closed-loop dynamics}
In analyzing the impact of the epistemic dimension on the closed-loop adoption dynamics induced by \eqref{eq:optimal_u} in 
a setting with unlimited resources, we first (informally) characterize the set of final adopters. In the unlimited resources scenario, solving \eqref{eq:optimal_problem} for a number of steps $t\rightarrow \infty$ to design policies nudging the acceptance of innovation leads to a progressive lowering of resistivity thresholds. In turn, this eventually helps adoption across the full network through social contagion.
Hence, differently from the case analyzed in Section \ref{sec:open_loop}, the full acceptance of the innovation is always asymptotically achieved, i.e., $S^{\star\star}=\mathcal{V}$.
Consequently, any effect induced by a credibility deficit or excess in this closed-loop scenario has not to be searched in the analysis of the final set $S^{\star\star}$, but rather on the \emph{transient} behavior of the diffusion process. We hence analyze it by first noticing that the gain shaping the control action is not influenced by the credibility deficit or excess (see \eqref{eq:controldesign_steps}). Instead, the epistemic dimension directly affects individual targets to which the input is proportional (see \eqref{eq:optimal_u}). In particular, based on \eqref{eq:individual_target}, the following holds:
\begin{align}\label{eq:equality_explain}
    \nonumber u_{x}(t) &=\kappa_{x}(t)\left[\rho_{x}^{u}(t)-\frac{\sum_{y \in N_{x}^{\star}(0)}{\gamma_{y,x}}}{\sum_{y \in N_x}{\gamma_{y,x}}}\right]
    =\underbrace{\kappa_{x}(t)(\rho_{x}^{u}(t)-1)}_{c_{1}(t)}+\kappa_{x}(t)\frac{\sum_{y \in (N_{x}^{\star}(0))^{c}}{\gamma_{y,x}}}{\sum_{y \in N_x}{\gamma_{y,x}}},
\end{align}
with $c_{1}(t)\leq 0$ due to the features of the control law for all $t$. 
The nudging input to an agent depends on the individual predisposition and is inversely proportional to the credibility of the agent's neighbors. This result implies that the higher impact adopters have (based on their credibility) on a non-adopter, the lower the input given to such an agent.
Moreover, given the dependence of $u_{x}(t)$ on $\rho_{x}^{u}(t)$, it is also easy to see that individuals with high resistance and low epistemic influence for neighboring adopters are the ones receiving more resources. Conversely, individuals with low resistance and high epistemic influence from neighboring adopters will receive minimal resources, as they are already inclined to adopt due to their individual attitude and mutual influence without the need for external incentives.
Based on this result and 
setting $N_{x}^{\alpha}=N_{x} \cap S^{\star}(0)$ and $N_{x}^{\beta}=N_{x} \cap (S^{\star}(0))^{c}$, we can now formalize the differences in input allocation and evolution between an epistemically fair and unfair closed-loop scenario as follows.  

\begin{lemma}\label{lemma:epistemic_infleunce_closedloop}
    Given an agent $x \in \mathcal{V}\setminus S^{\star}(t)$, assume that $\rho_{x}^{u}(t)$ and $a_{x}(t)$ to be the same for both the epistemically fair and unfair scenarios at a given time $t$. Then, credibility deficits/excesses lead to a reduction in the input given to the agent if
    \begin{equation}\label{eq:condition_clu}
       \Delta_x^{\beta}r^\alpha_x>\Delta_x^{\alpha}r^\beta_x,
    \end{equation}
    with $r_x^{\alpha}$, $r_x^{\beta}$, $\Delta_x^{\alpha}$ and $\Delta_x^{\beta}$ defined as in Proposition~\ref{cor:max_social_cont}, which, consequently, lessens the reduction of the agent's resistivity at time $t+1$.
\end{lemma}
\begin{proof}
    The proof can be found in the Appendix.
\end{proof}

\section{Numerical Examples in Closed-Loop}\label{sec:numerical_example}
By considering both an illustrative numerical example and a data-driven one, we empirically validate our formal results on the impact of epistemic bias in closed-loop. Both examples will lead to results that align with \cite{watts2002simple}, highlighting the double facet of credibility in innovation diffusion.\\

\begin{figure*}[t]
    \centering
    \begin{tabular}{ccc}
       \subfigure[Scenario 1\label{subfig:network1}]{ \includegraphics[width=0.3\linewidth]{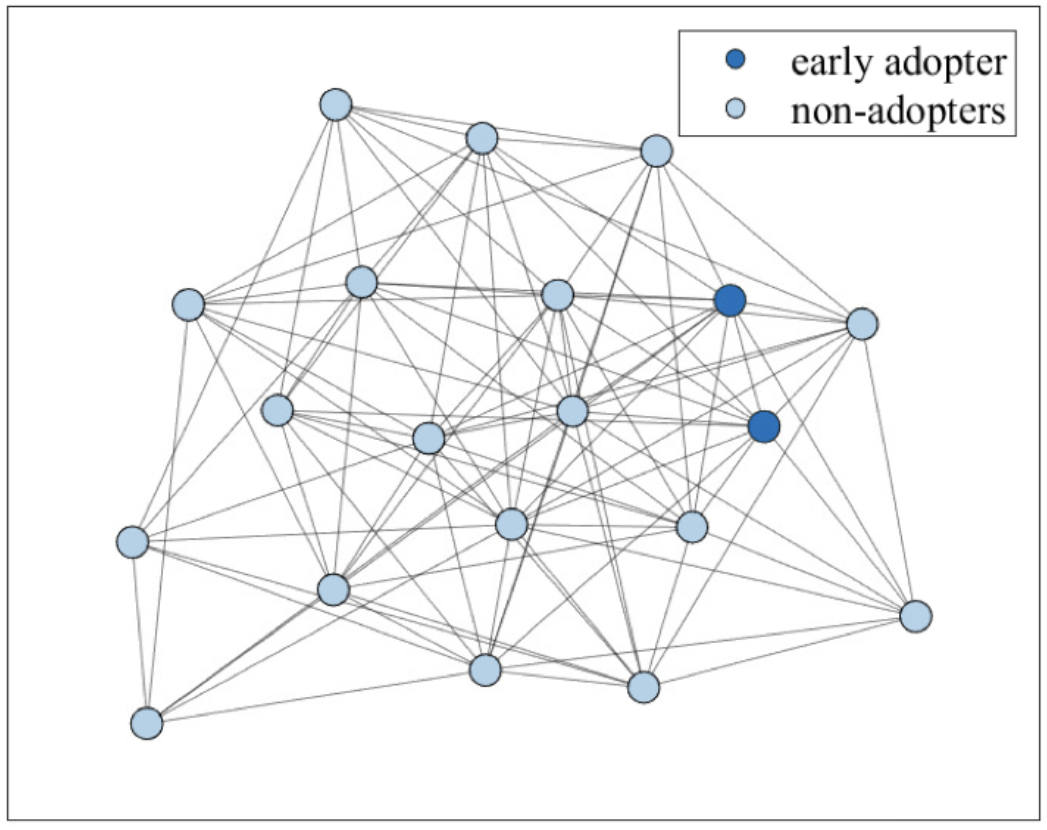}} & 
       \subfigure[Scenario 2\label{subfig:network2}]{ \includegraphics[width=0.3\linewidth]{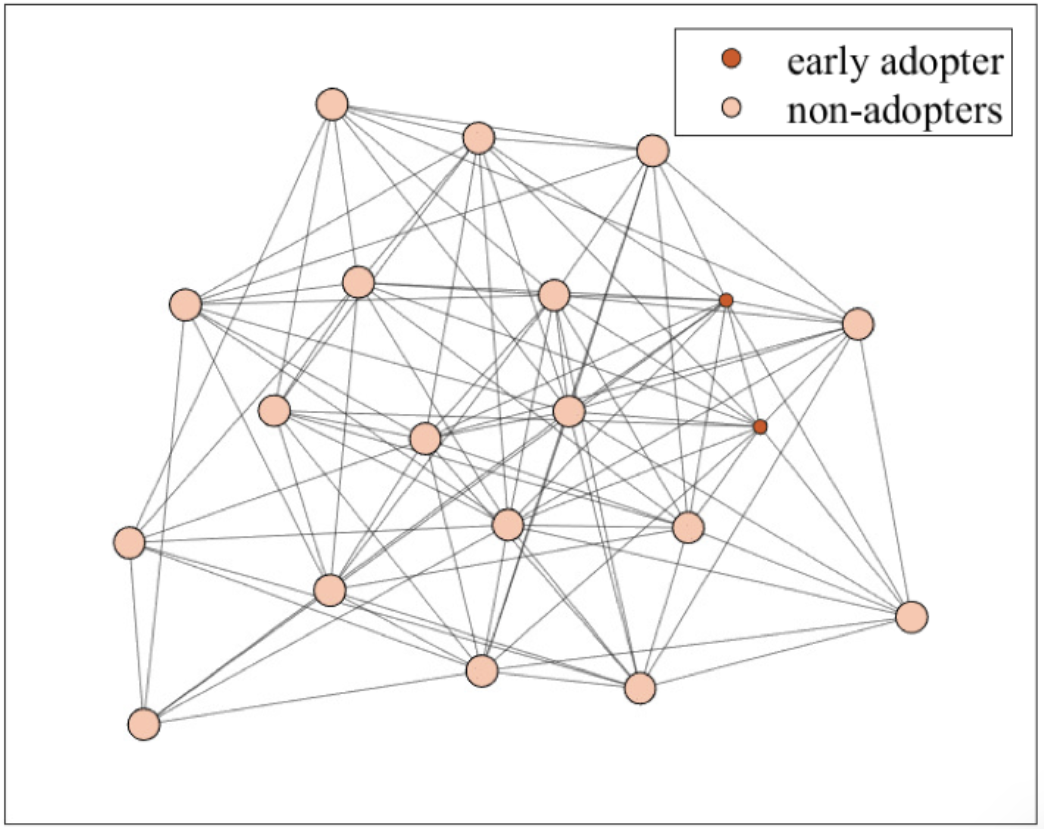}}
       &
       \subfigure[Scenario 3\label{subfig:network3}]{ \includegraphics[width=0.3\linewidth]{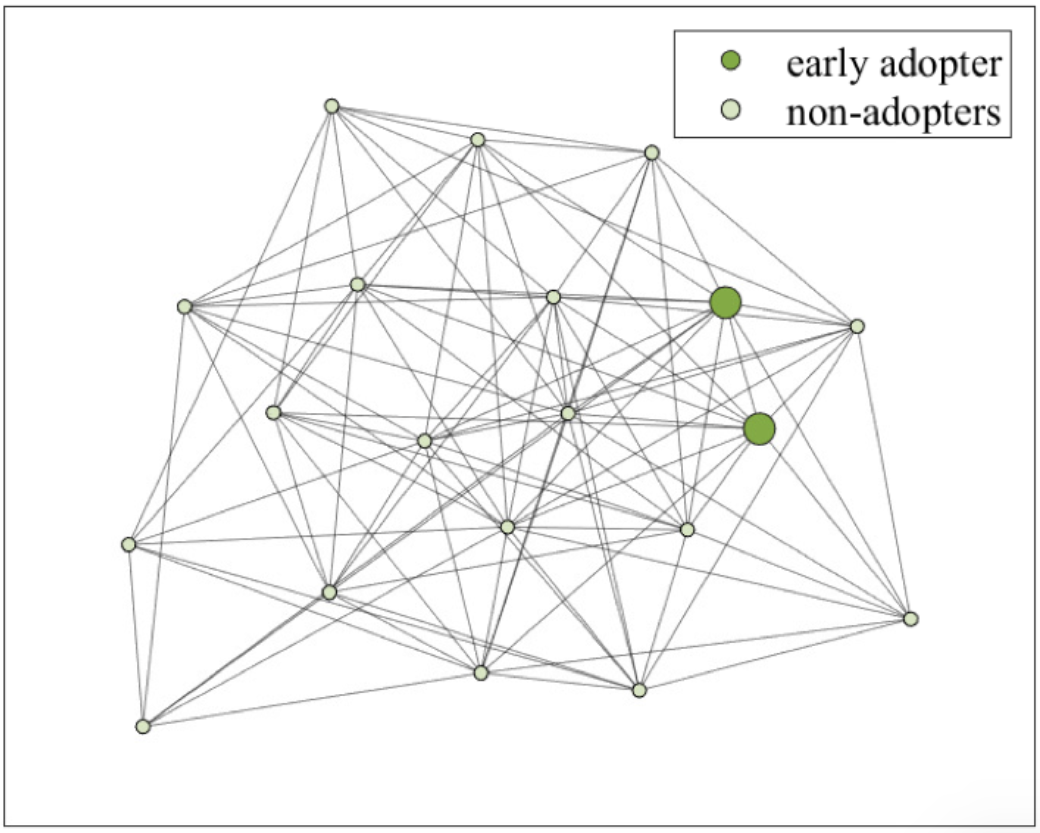}}
    \end{tabular}
    \caption{Comparative example: social network in the different epistemic scenarios, with node colors and sizes indicating agents' initial statuses and credibility.}
    \label{fig:network}
\end{figure*}

\noindent \textbf{Comparative example}
We first consider a population of $|\mathcal{V}|=20$ individuals, each interacting with a subset of randomly assigned agents\footnote{The probability that an edge between two agents exists is set to $0.5$.}. The agents in the seed set are also randomly selected, representing 10\% of the total population, i.e., $|S^{\star}(0)|=2$, while the others are assumed to verify $\rho_{x}^{u}(0)=0.8$, $b_{x}=r_{x}=1$ for all $x \in \mathcal{V}\setminus S^{\star}(0)$. For policy design, we further assume that $\bar{\omega}^{\rho}=\omega^{u}=1$, while setting $T=10$ (see \eqref{eq:optimal_problem}). Under these conditions, we compare three possible scenarios depicted in \figref{fig:network}: 1) same credibility deficit, i.e., $\gamma_{x,y}=\gamma_{y,x}$, $ \forall x,y \in \mathcal{V}$; 2) seeds are less credible than non-seeds, i.e., $\gamma_{x,y}<\gamma_{y,x}$, for all $x \in S^{\star}(0)$ and $y \in \mathcal{V}\setminus S^{\star}(0)$; 3) seeds are more credible than non-seeds, i.e., $\gamma_{x,y}>\gamma_{y,x}$, for all $x \in S^{\star}(0)$ and $y \in \mathcal{V}\setminus S^{\star}(0)$.
As all agents are equally (non-)credible, the first setting is ultimately an epistemically fair one, which is instead not the case for the other two.
\begin{table}[!t]
    \caption{Comparative example: performance indexes}
\label{tab:performace_policy}
    \centering
    \small
    \begin{tabular}{cccc}
         Scenario & $C $& $\bar{C}_x$ & $t^{**}$ \\
         \hline
         1& 12.5166 & 0.6954 & 7\\
         \hline
         2& 14.2320 & 0.7907 & 8\\
         \hline
         3& 6.2072 & 0.3448 & 6\\
         \hline
    \end{tabular}
\end{table}
The results attained in the three scenarios are quantitatively compared via the following indicators $C = \sum_{t=1}^{T}{\sum_{x \in \mathcal{V}}{u_x(t)}}$ and $\bar{C}_x = \frac{1}{N}\sum_{x \in \mathcal{V}}{\sum_{t=1}^{T}{u_x(t)}}$, denoting the total policy cost and the average individual policy cost, respectively.
The indicators are paired with the time $t^{**}$ required for full acceptance, i.e., $\{x \in \mathcal{V}| a_x(t^{**})=1\}\equiv\mathcal{V}$. 
As shown in Table \ref{tab:performace_policy}, our theoretical expectations are met. Comparing the first and the second scenarios, in the latter only the seeds are affected by a credibility deficit, impacting their influence on the community. This results in more resources being allocated to all individuals to reach full acceptance, increasing total and individual policy costs. The reduced credibility of seeds also affects the time required for the innovation to spread, despite nudging inputs. Oppositely, the third scenario reduces both costs and time to full acceptance compared to the nominal case, highlighting the potential positive effects of a credibility deficit in cascaded processes when it affects non-seeds only.

 \begin{figure}[!t]
    \centering
    \begin{tabular}{cc}
        \subfigure[Inputs \emph{vs} opposite epistemic weights $1-\bar{\rho}_x$ $t=0$. \label{subfig:inputs_t0}]{
            \includegraphics[width=0.4\columnwidth]{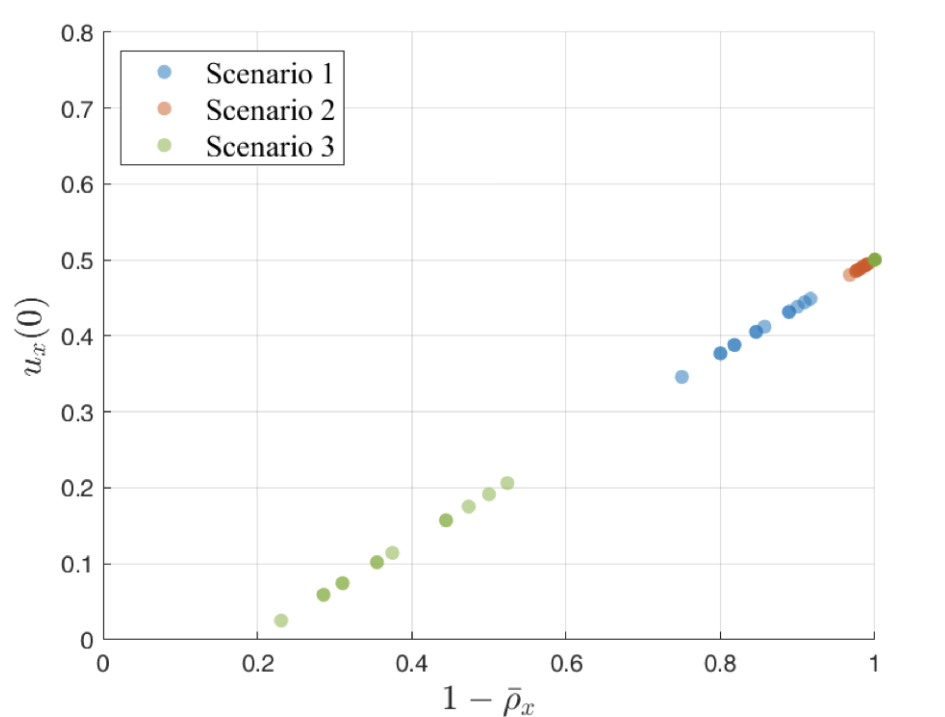}} &
        \subfigure[Thresholds \emph{vs} epistemic weights $\bar{\rho}_x$ at $t=1$.\label{subfig:thresholds}]{
            \includegraphics[width=0.4\columnwidth]{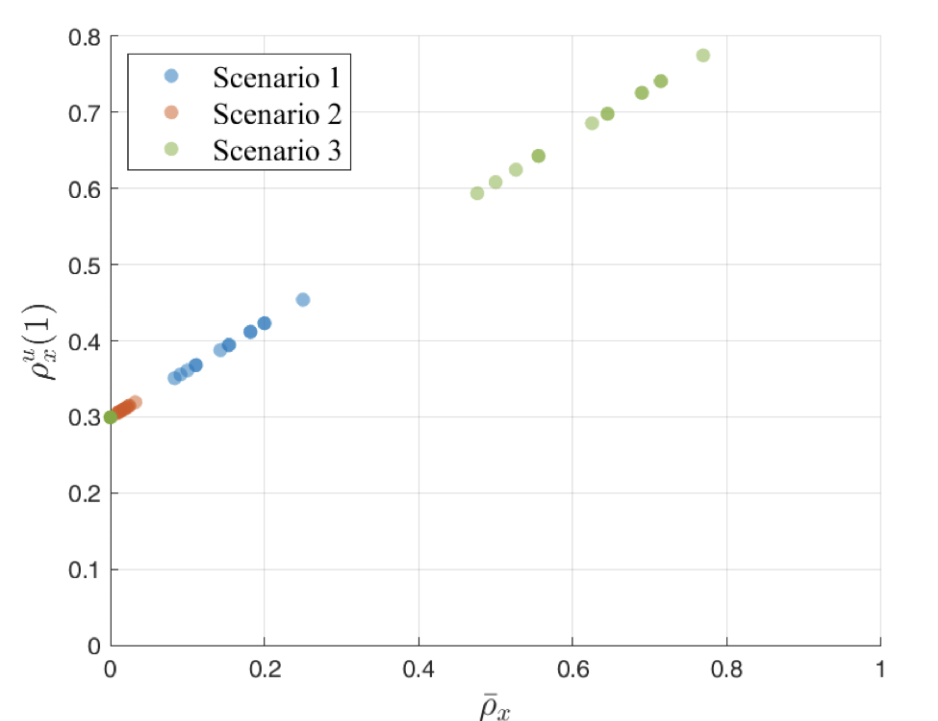}} \\
    \end{tabular}
    \caption{Comparative example: effect of the epistemic dimension on control actions and individual thresholds.}
    \label{fig:policy_simu}
\end{figure}

These effects are clear also when looking more in detail at the input at time $t=0$ and the consequent individual resistivities at time $t=1$, reported in \figref{fig:policy_simu} for all three scenarios.
Indeed, as shown in \figref{subfig:inputs_t0}, the input is clearly higher in the second scenario (when the influencing power of non-seed is amplified) than the third (when the influencing power of non-seed is reduced), with the nominal scenario placed in between. This behavior directly impacts the individual resistivity (see \figref{subfig:thresholds}), leading to a consistent reduction for those agents who are minimally affected by seeds in their proximity. Indeed, the diffusion of the innovation across the latter class of agents cannot benefit from the effect of social contagion, thus having to rely on a change of individual mindset. Conversely, agents more influenced by nearby adopters can embrace the innovation more easily through imitation.\\

\begin{figure}[!t]
    \centering
    \includegraphics[width=0.45\columnwidth]{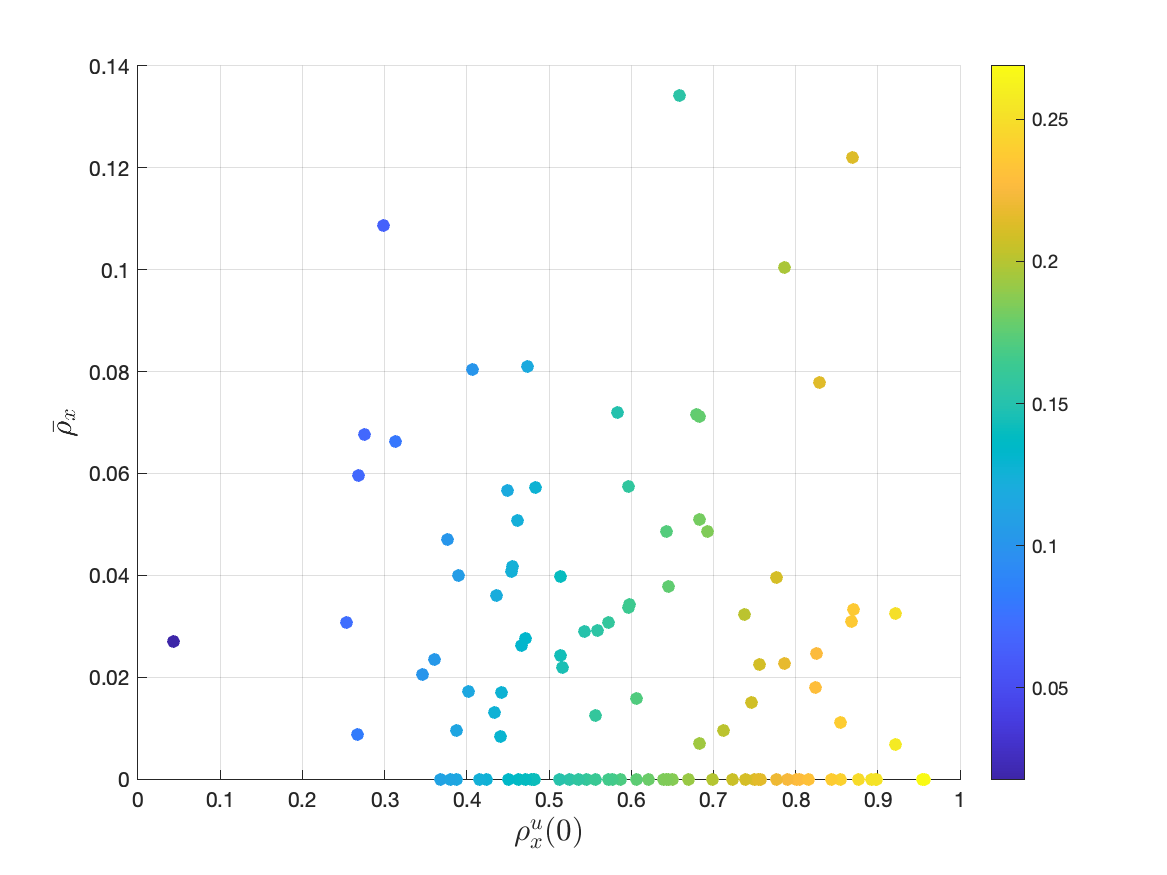}
    \caption{Data-driven scenario: control actions \emph{vs} relative epistemic weight.}\label{fig:actions_new}
\end{figure}

\noindent \textbf{Data-driven scenario.}
Inspired by the motivating examples in Section \ref{sec:motivation}, we leverage a data-driven approach to analyze the impact of the presented epistemic-based fostering policies in the diffusion of EV over the same prototypical population considered in \cite{villa2023fostering}. Starting from survey data~\cite{fiorello2015eu}, we consider network of $\mathcal{V}=168$ individuals 
, where the set of initial adopters $S^{\star}(0)$ comprises those respondents who declare already to possess an EV at the time of the survey. Following the methodology presented in \cite{villa2022sharing,villa2023fostering}, we quantify the resistivity $\rho_x(0)$ of each respondent $x$ in the survey to embrace the new mobility technology based on relevant socio-economic attributes (e.g., income level, age, environmental sensitivity). We consider the epistemic dimension of the policy design problem by computing individual reliability $\{r_x\}_{x \in \mathcal{V}}$ semi-randomly based on the educational level of the respondent $x$ 
. Moreover, through an intersectional effect, we account for the epistemic biases induced by the affiliation to specific minority groups by applying the computational strategy for relational credibility provided in Section \ref{sec:epistemic} (see Table \ref{tab:relational_cred}).
In this scenario, we find the optimal nudging policy solving \eqref{eq:optimal_problem} assigning equal importance to policy boosting effect and cost savings (i.e., imposing $\bar{\omega}^{\rho}=\omega^{u}=1$), and assuming for sake of simplicity that all agents equally react to policy actions\footnote{The reader is referred to \cite{villa2023fostering} for a possible data-driven definition of this parameter in the context of sharing mobility, here omitted to focus exclusively on epistemic-related aspects.} (i.e. $b_x=1\, \forall x\in\mathcal{V}$). 
Fusing on the initial time step, \figref{fig:actions_new} shows that optimal inputs tends to be greater for agents with higher resistivity and less influence from seed neighbors (bottom right corner of the graph). Conversely, inputs are minimal for agents with low resistivity and strong seed neighbor influence. These results support our theoretical insights, emphasizing the impact of individuals' epistemic capacity in the design of optimal nudging strategies, ultimately suggesting that incorporating epistemic considerations is essential for achieving fair policy design.

\begin{table}[]
    \centering
    \small
    \caption{Data-driven scenario: Example of data-driven computation of relational credibility.}
    \label{tab:relational_cred}
    \resizebox{0.4\columnwidth}{!}{
    \begin{tabular}{c c c c c c}
    \toprule
    \textbf{$x$} & \textbf{$y$} & \textbf{Educational Level} & \textbf{\(r_x\)} & \textbf{Shared Groups} & \textbf{\(\gamma_{x,y}\)} \\
    \midrule
    1 & 2 & High   & 0.8  & Gender, Income & 0.8 \\
    1 & 3 & High   & 0.8  & None           & 0.2 \\
    2 & 1 & Low    & 0.3  & Gender, Income & 0.15 \\
    2 & 3 & Low    & 0.3  & Age            & 0.075 \\
    3 & 1 & Medium & 0.6  & None           & 0.3 \\
    3 & 2 & Medium & 0.6  & Age            & 0.6 \\
    \bottomrule
    \end{tabular}}
\end{table}

\section{Conclusions}\label{sec:concl}
By merging 
theoretical conceptualizations and a quantitative approach, in this work we have extended the conceptualization of algorithmic fairness in an epistemic direction, focusing on the context of automated tools for policy design with a focus on innovation diffusion. We equipped a (standard) irreversible cascade dynamical model with an epistemic characterization, modeling the 
credibility of agents, to 
lay the foundations for new control-oriented approaches for an epistemically fair policy design. 
The present work shows the importance of an epistemic extension of algorithmic fairness, in particular for recognizing and addressing a distinct type of harm which is mostly neglected in the current debate. Our empirical study evidences that any nudging strategy neglecting the epistemic dimension risks reinforcing existing biases, leading to unfair outcomes. The quantitative analysis we provided ultimately allows to highlight the pivotal role of the epistemic dimension in shaping tools for the innovation diffusion dynamics and policy design. 
Future research will 
involve an epistemic-oriented investigation of more complex models for adoption dynamics (e.g., \cite{916272ae-6b7a-3e62-b7e2-5747187dae7b}) to introduce a stochastic component for a more realistic representation of the diffusion process. 

\appendix

\section*{Proof of Corollary~\ref{cor:max_social_cont}}
For the case with a credibility deficit/excess to result in (at least) the same set of positively inclined agents at time $t+1$ as the epistemically fair scenario, it has to hold that
\begin{equation*}
    \frac{\sum_{y \in N_{x}^{\alpha}} \gamma_{y,x}}{\sum_{y \in N_{x}} \gamma_{y,x}}\geq \frac{\sum_{y \in N_{x}^{\alpha}} r_{y}}{\sum_{y \in N_{x}} r_{y}},
\end{equation*}
at least for all $x \in {S}^{\star,\mathrm{fair}}(t+1)$. In turn, 
this inequality holds if $\left(r^{\alpha}_{x}-\Delta ^{\alpha}_{x}\right)r^{\beta}_{x}-r^{\alpha}_{x}(r^{\beta}_{x}-\Delta^{\beta}_{x}) \geq 0$, at least for all $x \in {S}^{\star,\mathrm{fair}}(t+1)$. Simple manipulations of this inequality lead to \eqref{eq:condition_transient}, thus concluding the proof.

\section*{Proof of Lemma~\ref{lemma:epistemic_infleunce_closedloop}}
Our result straightforwardly follows from the definitions in \eqref{eq:optimal_u}, as from it we get
\begin{align*}
    u_{x}(t)&=
    \kappa_{x}(t)\left(\rho_{x}^{u}(t)-\frac{r_{x}^{\alpha}}{r_{x}^{\alpha}+r_{x}^{\beta}}\right) +\kappa_{x}(t)\left(\frac{r_{x}^{\alpha}}{r_{x}^{\alpha}+r_{x}^{\beta}}-\frac{r_{x}^{\alpha}-\Delta_{x}^{\alpha}}{r_{x}^{\alpha}-\Delta_{x}^{\alpha}+r_{x}^{\beta}-\Delta_{x}^{\beta}}\right)
\end{align*}
where the first term is the nudging input given to agent $x$ at time $t$. Since $\kappa_{x}(t)\geq0$, for the credibility deficit or excess to reduce the input given to the system, the following has to hold
\begin{align*}
\frac{r_{x}^{\alpha}\left(r_{x}^{\alpha}-\Delta_{x}^{\alpha}+r_{x}^{\beta}-\Delta_{x}^{\beta}\right)-\left(r_{x}^{\alpha}-\Delta_{x}^{\alpha}\right)(r_{x}^{\alpha}+r_{x}^{\beta})}{(r_{x}^{\alpha}+r_{x}^{\beta})\left(r_{x}^{\alpha}-\Delta_{x}^{\alpha}+r_{x}^{\beta}-\Delta_{x}^{\beta}\right)}<0.
\end{align*}
Being the denominator positive, a reduction in the input happens when
\begin{equation*}
    r_{x}^{\alpha}\left(r_{x}^{\alpha}-\Delta_{x}^{\alpha}+r_{x}^{\beta}-\Delta_{x}^{\beta}\right)-\left(r_{x}^{\alpha}-\Delta_{x}^{\alpha}\right)(r_{x}^{\alpha}+r_{x}^{\beta})< 0,
\end{equation*}
which, after straightforward manipulations, leads to the condition in \eqref{eq:condition_clu}. By plugging the expression for $u_{x}(t)$ obtained previously into \eqref{eq:dynamic}, the thesis easily follows.


\bibliographystyle{ACM-Reference-Format}
\bibliography{references}


\end{document}